# NMR studies of Successive Phase Transitions in $Na_{0.5}CoO_2$ and $K_{0.5}CoO_2$

Mai Yokoi, Yoshiaki Kobayashi, Taketo Moyoshi and Masatoshi Sato

*Department of Physics, Division of Material Science, Nagoya University,
Furo-cho, Chikusa-ku, Nagoya 464-8602*

$^{59}$Co- and $^{23}$Na-NMR measurements have been carried out on polycrystalline and *c*-axis aligned samples of $Na_{0.5}CoO_2$, which exhibits successive transitions at temperatures $T = 87$ K ($= T_{c1}$) and $T = 53$ K ($= T_{c2}$). $^{59}$Co-NMR has also been carried out on *c*-axis aligned crystallites of $K_{0.5}CoO_2$ with similar successive transitions at $T_{c1} \sim 60$ K and $T_{c2} \sim 20$ K. For $Na_{0.5}CoO_2$, two sets of three NMR lines of $^{23}$Na nuclei explained by considering the quadrupolar frequencies $\nu_Q \sim 1.32$ and 1.40 MHz have been observed above $T_{c1}$, as is expected from the crystalline structure. Rather complicated but characteristic variation of the $^{23}$Na-NMR spectra has been observed with varying $T$ through the transition temperatures, and the internal fields at two crystallographically distinct Na sites are discussed on the basis of the magnetic structures reported previously. The internal fields at two distinct Co sites observed below $T_{c1}$ and the $^{59}1/T_1$-$T$ curves of $Na_{0.5}CoO_2$ and $K_{0.5}CoO_2$ are also discussed in a comparative way.

## 1. Introduction

$Na_xCoO_2$ with triangular lattice of edge-sharing $CoO_6$ octahedra is the mother system of the superconducting $Na_{0.3}CoO_2 \cdot 1.3H_2O$.[1] For this system, characteristics of the *T*-dependences of magnetic susceptibility $\chi$ and the *x*-dependence of the electronic specific heat coefficient $\gamma$ have been reported to be distinct between two regions separated by the boundary at $x \sim 0.6$.[2,3] At $x \cong 0.5$, successive phase transitions take place at $T_{c1} \sim 87$ K and $T_{c2} \sim 53$ K.[2] The susceptibility $\chi$ is suppressed by the transitions with decreasing *T*, except in the case of the transition at $T_{c1}$ in the magnetic field $H \parallel c$. An anomaly of the electrical resistivity $\rho$ observed at $T_{c1}$ is rather small, while it exhibits a sharp increase at $T_{c2}$ with decreasing $T$ (see Fig. 1(a)).[2]

We have been studying the successive phase transitions by various experimental methods including NMR/NQR and neutron measurements[3] to extract information on the electronic state of $Na_xCoO_2$ and revealed that an antiferromagnetic state is realized below $T_{c1}$. There are two crystallographically distinct Co sites, Co1 and Co2 sites with different ordered moments. Co1 sites have antiferromagnetic in-plane ordering with their directions within the *ab*-plane and the moments at Co2 sites align antiferromagnetically with their directions parallel to *c*, though we could not distinguish if the in-plane ordering of Co2 moments is ferromagnetic or antiferromagnetic. The magnitudes of the ordered moments at Co1 and Co2 sites are $\sim 0.3$ and $\sim 0.1$ $\mu_B$ at 8 K. However, the ordered moments at Co2 sites were not detected by polarized neutron measurements,[4] which is possibly due to a slight difference between the samples used in these measurements.

Significant anomalies due to the transition at $T_{c2}$ have been observed in the data of various quantities such as $\rho$ and $\chi$.[2,3,5,6] Moreover, the gap-opening has been observed by optical measurements[7] and angle-resolved photoemission spectroscopy (ARPES).[8] However, as reported in our previous paper, no significant anomalies have been observed at the temperature in the NMR spectra for both Co sites, except an appearance of the very small splitting of the zero-field NMR peaks from Co2 sites with decreasing $T$.[3] No significant anomaly has been found in the *T*-dependence of the magnitudes of the Co1 moments, either. In order to clarify what anomalies exist in the NMR data, we have carried out further detailed NMR studies.

Theoretically, several ideas have been proposed as the origin of the transition at $T_{c2}$. A charge-disproportionation,[9] an antiferromagnetic order on the orthorhombic superlattice of triangular lattice,[10] the Na ion ordering[11] and the in-plane ferromagnetic ordering of Co2 moments at $T_{c2}$[12] are the examples.

In this paper, we present results of $^{59}$Co-NMR and $^{23}$Na-NMR measurements focussing on the *T* region around $T_{c2}$, and also present results of comparative studies of $K_{0.5}CoO_2$, which exhibits similar successive transitions at $T_{c1} \sim 60$ K and $T_{c2} \sim 20$ K.[8,13,14]

## 2. Experiments

Polycrystalline samples of $Na_xCoO_2$ ($x \sim 0.75$) were prepared at 750-850°C by solid reaction for 24 h, and single crystals, which also have the *x* value of $\sim 0.75$ were prepared by the floating zone (FZ) method. Then, these samples were immersed in the $Br_2/CH_3CN$ solution for several days to de-intercalate Na ions, where the concentration of the solution was controlled to obtain the samples witth $x \sim 0.5$. The obtained samples were found to be the single phase, and by using the lattice parameter *c*-*x* curve,[2] the *x* values of the used samples were determined to be 0.5±0.02.

Single crystals of $K_{0.5}CoO_2$ were grown by the FZ method: Mixtures of KOH and $Co_3O_4$ with the molar ratio K:Co = 1:1 were preheated at 400°C and pressed into rods for the use of the crystal growth. Obtained crystals were washed with $C_2H_5OH$ to remove $K_2CO_3$. These crystals were found to have the K-concentration of about 0.5 by the X-ray Rietveld analyses.[14]

The magnetic susceptibility $\chi$ was measured by Quantum Design SQUID magnetometer. The electrical resistivity $\rho$ was measured by the four terminal method with increasing $T$.

The $^{59}$Co-NQR/NMR and $^{23}$Na-NMR measurements were carried out by the standard coherent pulse method. The $^{59}$Co-NMR/NQR (or $^{23}$Na-NMR) spectra were measured by recording the nuclear spin-echo intensities $I$ with the applied frequency or magnetic field being changed stepwise. The $^{59}$Co longitudinal nuclear relaxation rates $1/T_1$ were obtained by measuring the intensity as a function of the time elapsed after the saturation pulses.

## 3. Experimental Results

NMR/NQR spectra of $Na_{0.5}CoO_2$ indicate that above $T_{c1}$ $Na_{0.5}CoO_2$ has two Co sites, Co1 and Co2 sites. (The assignment of these sites can be found in the previous paper.[3]) Their quadrupole frequencies $\nu_Q$ are estimated to be 4.1 MHz and 2.8 MHz, respectively at 90 K ($> T_{c1} \sim 87$ K). $K_{0.5}CoO_2$ also has two Co sites and the $\nu_Q$ values of Co1 and Co2 sites are estimated at 80 K ($> T_{c1} \sim 60$ K) to be 3.8 and 3.1 MHz, respectively.[14] Below $T_{c1}$, we have also measured the zero-field NMR spectra of $^{59}$Co. In Figs. 1(b) and 1(c), the spectra observed for $Na_{0.5}CoO_2$ (powder) at 9.8 K ($< T_{c2}$) and for $K_{0.5}CoO_2$ (single crystal) at 4.2 K ($< T_{c2}$), respectively,[3,14] are recited together with the positions and intensities of the NMR peaks calculated by the perturbation of the $eqQ$ interaction up to the second order (middle) and by the exact diagonalization (top), where the black and grey lines in the top and middle panels correspond to the contributions to Co1 and Co2 sites, respectively. For both Co1 and Co2 sites, the main contribution of the 1/2 ↔ -1/2 transitions are found in the lines indicated by the open circles and open squares, respectively.

For $Na_{0.5}CoO_2$, the lines of Co2 sites exhibit the splitting as already reported in ref. 3 below $T_{c2}$. We define the sites with the larger resonance frequency as Co2(a) and another sites as Co2(b). For $K_{0.5}CoO_2$, no indication of the splitting can be found, as shown in Figs. 2(a)-2(c), even as the change of the line broadening.

In the perturbation calculation for $Na_{0.5}CoO_2$, we obtained the following parameters at 9.8 K: For Co2(a), the anisotropic parameter $\eta = 0.25$, $\nu_Q (= \nu_{zz}) = 2.82$ MHz, the internal field $H_{int} = 0.693$ T and $\theta = 0$ ($\nu_{zz} > \nu_{xx} > \nu_{yy}$; $z \parallel c^3$). For Co2(b), $\eta = 0.25$, $\nu_Q = 2.835$ MHz, $H_{int} = 0.673$ T and $\theta = 0$ ($z \parallel c^3$). For Co1, $\eta = 0.4$, $\nu_Q = 4.05$ MHz, $H_{int} = 2.077$ T, $\theta = 90°$ and $\varphi = 0$ ($z \parallel c^3$).

The spin directions of Co1 and Co2 sites below $T_{c1}$ were found in our previous paper[3,14] to be, perpendicular and parallel to $c$, respectively, from the zero-field NMR spectra. It was also found that the principal direction $z$ of the quadrupole tensor is parallel to $c$. Then, the Co2 moments is parallel to $z$ ($\parallel c$), and $\nu_Q$ ($= \nu_{zz}$) of Co2 sites can be easily determined. It does not change with $T$ through $T_{c1}$.[3] We have first tried to calculate the spectra for Co2 sites by using the values of $\nu_Q$ and $H_{int}$ roughly estimated by considering the NMR spectra obtained in the external magnetic field $H$ in our previous study.[3] We have also used $\theta = 0$ and the $\eta$ value estimated roughly by using the powder pattern of the Co NMR. For these parameters, we have found the observed spectra can be reproduced reasonably well. Then, we have finally fitted the calculated spectra to the observed data by using $\nu_Q$ and $H_{int}$ as the fitting parameters with the other parameters being fixed, and obtained the above parameters. We note that $\nu_Q$ does not change significantly with varying $T$.

For Co1 sites, because the spin direction of Co1 is not parallel to $z$, there is an ambiguity in the estimation of $\nu_Q$. However, if we assume that $\nu_Q$ remains unchanged through $T_{c1}$ as in the case of Co2 sites, we can estimate the values of $\theta$, $\varphi$ and $\eta$ for the $H_{int}$ value determined from the data obtained in the field ($\parallel ab$-plane) at each temperature. By this process, we find $\theta \sim 90°$ and $\varphi \sim 0$, indicating that the $x$ direction is parallel to the spin direction ($\parallel \boldsymbol{b}$ as shown by neutron diffraction data[3]). Because the direction with $\theta \sim 90°$ and $\varphi \sim 0$ is naturally expected for the present system, we fix these values at $\theta = 90°$ and $\varphi = 0$ in the following calculations. Then, we have carried out the fitting of the spectra to the observed data with $\nu_Q$, $\eta$ and $H_{int}$ as the fitting parameters, and obtained the values described above. Here, we note that $H_{int}$ is almost independent of the choice of $\nu_Q$ and $\eta$. It is also true that $\nu_Q$ does not sensitively depend on $T$ through $T_{c2}$.

The present assignment is different from the erroneously reported assignment of ref. 3, although this correction was already shown in ref. 14. On this point, we have checked the correctness of the present assignment by studying the recovery curves of the longitudinal $^{59}$Co nuclear magnetization at the peak positions. (The difference among the detailed $m$-dependence of $m \leftrightarrow m+1$ transition curves enables us to make this confirmation.) Based on this result, the internal magnetic fields $H_{int}$ and $^{59}$Co relaxation rates $1/T_1$ are re-estimated as shown later.

For $K_{0.5}CoO_2$, the similar calculations to those for $Na_{0.5}CoO_2$ have been carried out and the obtained parameters are as follows. For Co2, $\eta = 0.2$, $\nu_Q = 3.15$ MHz, $H_{int} = 0.582$ T and $\theta = 0$ ($z \parallel c^3$). For Co1, $\eta = 0.7$, $\nu_Q = 3.82$ MHz, $H_{int} = 1.29$ T, $\theta = 90°$ and $\varphi = 30°$ ($z \parallel c^3$). In this case, because the $\nu_Q$ value cannot be considered to be much smaller than the nuclear Zeeman energy, the perturbation calculation does not present satisfactory results. However, the values of $\nu_Q$ and $H_{int}$ do not exhibit a significant deviation from the corresponding values determined by the exact diagonalization method. (On the assignments carried out by the exact diagonalizations, we added a note in the last part of the present paper.)

In Figs. 3(a) and 3(b), the values of $H_{int}$ at both Co1 and Co2 (Co2(a) and Co2(b) for $Na_{0.5}CoO_2$ below $T_{c2}$) sites are plotted against $T$ for $Na_{0.5}CoO_2$ and $K_{0.5}CoO_2$, respectively. They have been obtained by fitting the peak positions of the perturbation calculations up to the second order to those of the observed ones with $H_{int}$ and $\nu_Q$ being the fitting parameters and with other parameters being fixed. The $\nu_Q$ values have been found not to depend on $T$ sensitively. (The square points show the results of the exact diagonalizations and explained later.)

The static magnetic moments appear at $T_{c1}$ at both of Co1 and Co2 sites. The ratios of $H_{int}$ values of Co1 and Co2 sites are ~ 3:1 and ~ 2:1 at low $T$ for $Na_{0.5}CoO_2$ and $K_{0.5}CoO_2$, respectively. (Because the $H_{int}$ values at Co2(a) and Co2(b) sites of $Na_{0.5}CoO_2$ below $T_{c2}$ are almost equal, we do not distinguish Co2(a) and Co2(b) in the calculation of the ratios.) For $K_{0.5}CoO_2$, $H_{int}$ at Co1 sites exhibits a slight anomaly at $T_{c2}$, while $H_{int}$ at Co2 sites becomes nearly $T$-independent below $T_{c2}$ as $T$ decreases through $T_{c2}$ (see the inset).

The very small splitting of $H_{int}$ at Co2 sites of $Na_{0.5}CoO_2$, which begins at $T_{c2}$ with decreasing $T$ may be one of key elements for understanding the mechanism of the transition at $T_{c2}$. However, we note here that such kind of splitting does not exist in $K_{0.5}CoO_2$, which makes it unclear whether the splitting commonly characterizes the transition (see Figs. 2(a) and 2(b)).

Figure 4 shows the $T$-dependence of the longitudinal relaxation rate divided by $T$, $1/T_1T$ of $^{59}Co$ of $K_{0.5}CoO_2$ together with that of $Na_{0.5}CoO_2$[3]. The data of $Na_{0.5}CoO_2$ at Co2 sites below $T_{c1}$ are the results of the re-analyses of the first satellite peak of new assignment. Above $T_{c1}$, the data of $K_{0.5}CoO_2$ were taken at the frequency $\nu_{Q3}$ ($\cong 3\nu_Q$) of the NQR spectra of the $\pm 5/2 \leftrightarrow \pm 7/2$ transition for both Co1 and Co2 sites, and below $T_{c1}$, they were taken at the peak positions of the center transition lines and third satellites of the zero-field NMR spectra of Co1 and Co2 sites, respectively.

For $Na_{0.5}CoO_2$, the data above $T_{c1}$ were taken at the frequency $\nu_{Q3}$ of the NQR spectra for both Co1 and Co2 sites. Because the frequency $\nu_{Q3}$ of Co2 sites is nearly equal to that of the $\pm 3/2 \leftrightarrow \pm 5/2$ transition ($\equiv \nu_{Q2}$) at Co1 sites, we have analyzed, to deduce the $1/T_1$ of Co2 sites, the recovery curves observed at the frequency $\nu_{Q3}$ of Co2, assuming that two contributions of the Co1-$\nu_{Q2}$ and Co2-$\nu_{Q3}$ signals coexist. Then, we have found from the difference between the shapes of the curves that the latter contribution is predominant, which guarantees that we have safely obtained the $1/T_1$ values of Co2. (In the analysis, we have also used the $1/T_1$ values of Co1 sites already observed at the $\nu_{Q3}$ position.) In the measurements of $1/T_1$ below $T_{c1}$, the peak positions of first satellite lines and third satellites of the zero-field NMR spectra were used at Co1 and Co2 sites, respectively. The $1/T_1T$ values of Co1 sites below $T_{c1}$ become a little larger than that reported in ref. 3 after the re-estimation.

At $T_{c1}$, sharp peaks of the $1/T_1T$-$T$ curves have been observed at all sites, as our previous data of $Na_{0.5}CoO_2$ showed.[3] At $T_{c2}$, $1/T_1T$-$T$ curves of both Co sites also exhibit sharp peaks for $K_{0.5}CoO_2$ in a clear contrast to the case of $Na_{0.5}CoO_2$, where only the rapid decrease of $1/T_1T$ is observed at $T_{c2}$ with decreasing $T$. The magnetic nature of the transition seems to appear more strongly in $K_{0.5}CoO_2$ than in $Na_{0.5}CoO_2$. The ratios of $1/T_1T$ values of Co1 and Co2 sites at 200 K ($> T_{c1}$) are 3:1 and 2:1 for $Na_{0.5}CoO_2$ and $K_{0.5}CoO_2$, respectively.

$^{23}Na$-NMR measurements have also been carried out using a sample of $c$-axis aligned crystallites of $Na_{0.5}CoO_2$. The NMR spectra obtained by varying frequency in the fixed field ($H \sim 4.1$ T) parallel to the $c$-axis are shown in Fig. 5 at various temperatures. Above $T_{c1}$, the spectra consist of the contributions from two crystallographically distinct Na sites, and each contribution has a set of three lines split by the electric quadrupole interaction as labeled A and B in Fig. 5 at 97.6 K. The integrated intensity of the spectra from A sites is almost equal to that of B as is expected from the crystal structure. The $\nu_Q$ values of A and B sites are estimated to be 1.40 MHz and 1.32 MHz at 97.6 K, respectively. Note that there are two lines at the center position with nearly equal Knight shifts, indicating that the $^{23}Na$ nuclei at both sites see almost equal hyperfine fields.

Below $T_{c1}$, the lines from the A and B sites shift to both the lower and higher frequency sides due to the internal magnetic field created by the ordered moments at Co sites. As the result, the set A splits into two sets $\alpha$ and $\alpha'$, and the set B splits into two pairs of sets, $\beta_1$ and $\beta_1'$, and $\beta_2$ and $\beta_2'$ (see Fig. 5). These sets, each of which consists of three lines, can be distinguished by the $\nu_Q$ values. The integrated intensities of the spectra of $\beta_2$ and $\beta_2'$ sets is rather small (about 10% of whole spectra). If these weak spectra are due to certain kinds of regular modulation of the magnetic moments at Co1 and/or Co2 sites, we should observe spectral changes (splittings) at these sites, too. Because we have not observed these changes, the spectra are considered to stem from certain lattice imperfections. The $\nu_Q$ values of these sites do not change significantly with $T$ at $T_{c1}$ and $T_{c2}$, but gradually decrease with decreasing $T$, and at 5 K, $\nu_Q$ values are estimated to be 1.38 MHz ($\alpha$), 1.31 MHz ($\beta_1$) and 1.32 MHz ($\beta_2$). (The sites $\alpha$, $\beta_1$ and $\beta_2$ represent the pairs of the sets $\alpha$ and $\alpha'$, $\beta_1$ and $\beta_1'$, and $\beta_2$ and $\beta_2'$, respectively.)

Figure 6 shows the $T$-dependence of the parallel component $H_{int}^c$ of the internal field $H_{int}$ to the applied magnetic field (parallel to $c$, here). Each value of $H_{int}^c$ was estimated approximately by considering the relation $H \gg H_{int}$, as the half distances between the center lines of each pair of sets $\alpha$ and $\alpha'$, $\beta_1$ and $\beta_1'$, and $\beta_2$ and $\beta_2'$. The field $H_{int}^c$ appears at $T_{c1}$ at every Na site. The magnitudes of $H_{int}^c$ at these sites at 5 K are estimated to be 166 G ($\alpha$), 691 G ($\beta_1$) and 435 G ($\beta_2$), respectively. Although anomalies exist at $T_{c2}$ at least in the $H_{int}^c$-$T$ curves of the $\beta_1$ and $\beta_2$ pairs, they can hardly be distinguished in the figure. Below $T_{c1}$, there does not exist appreciable intensity of $^{23}Na$-NMR lines at the position corresponding to $H_{int}^c = 0$ (see Fig. 5), which is in a clear contrast to the data reported in ref. 15.

The internal fields at Na sites are induced by the ordered Co moments via the transferred hyperfine interaction and dipole fields. Because our neutron structure analyses have revealed that the Na position deviates from the midpoint between two adjacent $CoO_2$ layers,[16] Na atoms can feel the transferred hyperfine field but only along *c*. The Co1 moments do not contribute to the transferred hyperfine fields at Na sites, because they are always two antiparallel spins, whose contributions are cancelled. The magnitudes of the dipole fields along two directions indicated in Fig. 7 at Na2 sites are shown in the right panel of Fig. 7. They are roughly estimated by considering just the nearest neighbor Co atoms. The field along *c* created by the Co1 moments $\mu_1 \sim 0.34\ \mu_B$[3] is about 400 G. The field within the *c* plane created by the Co2 moments $\mu_2 \sim \mu_1/3$ is $\sim 80$ G. In these calculations, we have not considered the deviation of Na atoms stated above.[16] At Na1 sites, the dipole fields from the two nearest neighbor Co2 sites is zero (see left panel of Fig. 7), and the fields along and perpendicular to *c* are $\sim 20$ G and $\sim 5$ G, respectively, even when the contributions from the twelve next nearest neighbor Co sites in the figure are considered. Therefore, the main origin of $H_{int}$ may be the transferred hyperfine interaction. Here, comparing the calculated internal fields $H_{int}^c$ with the observed values, 166 G at α sites and 691 G at $\beta_1$ sites, we tentatively assign that α and $\beta_1$ sites correspond to Na1 and Na2 sites, respectively.

Figure 8 shows the $^{23}$Na-NMR spectra taken at two temperatures for the *c*-axis aligned sample of Na$_{0.5}$CoO$_2$. The data at 98.8 K (closed circles) were taken by changing the NMR frequency in the fixed field $H \sim 4.1$ T parallel to the *ab*-plane, and the data at 8.1 K (open circles) were taken by changing the applied field *H* (|| *ab*-plane) with the frequency *f* being fixed. As we have already stated, there exist two crystallographically distinct Na sites A and B even above $T_{c1}$. The NMR spectra at 98.8 K ($> T_{c1}$) can be explained by using the $\nu_Q$ values of 1.32 and 1.26 MHz and by the η values of 0.62 and 0.52 for A and B sites, respectively (see the lower panel of Fig. 8). The calculated line for these parameters obtained by considering the *eqQ* interaction up to the second order is shown by the solid line in the lower panel. Note that these $\nu_Q$ values are about 5 % smaller than those estimated from the spectra for *H* || *c* at 97.6 K ($> T_{c1}$). The discrepancy of $\nu_Q$ values may be caused by their sample misalignment. Below $T_{c1}$, the existence of two components with sharp and broad linewidths can be distinguished in the central line spectra, where the sharper one has the double peak structure as is shown in the upper panel of Fig. 8. These NMR spectra can essentially be explained by the in-plane components of internal fields $H_{int}^{ab}$, where the magnitudes of $H_{int}^{ab}$ are $\sim 40$ G ($\nu_Q = 1.30$ MHz, η = 0.62) and $\sim 120$ G ($\nu_Q = 1.25$ MHz, η = 0.47) at 8.1 K ($< T_{c2}$) for α and $\beta_1$ (or $\beta_2$) sites, respectively. The calculated line is shown by the solid line in the upper panel. In this calculation, we have considered the first order *eqQ* interaction.

Although the observed values of $H_{int}^{ab}$ ($\sim 120$ G) and $H_{int}^c$ ($\sim 691$ G) at $\beta_1$ sites are 1.5-1.7 times larger than the calculated values, it may be attributed to the fact that the calculated values are obtained by a simple model that only the dipole fields from the moments without spreading and/or within the narrow region near the relevant $^{23}$Na position are considered. We think that the result indicates that the dipole field predominantly contributes to $H_{int}$ at Na2 sites.

In Fig. 9, the *T*-dependence of $^{23}$Na-NMR Knight shift *K* ($> T_{c1}$) or $K_m$ ($< T_{c1}$) of Na$_{0.5}$CoO$_2$ is compared with that of χ in the field parallel to *c*, where $K_m$ is defined below $T_{c1}$ as the average peak positions of each pair of the central transition lines of α and α', $\beta_1$ and $\beta_1$', and $\beta_2$ and $\beta_2$' sets. The hyperfine coupling constant $A_{hf}$ is estimated to be 6.7 kOe/$\mu_B$ from the *K*-χ plot in the *T* region between 100 to 260 K. The orbital part of the susceptibility is estimated to be $1.1 \times 10^{-4}$ emu/mol, which is nearly equal to the value ($0.79 \times 10^{-4}$ emu/mol) observed by $^{59}$Co-NMR measurements for *H* || *c*. *K* decreases gradually with decreasing *T* from room temperature to $T_{c1}$. At $T_{c1}$, *K* ($K_m$) of α sites begins to increase and those of $\beta_1$ and $\beta_2$ sites begin to decrease. Because the internal fields at these Na sites created by the Co moments are cancelled by taking the average positions of the two split peaks of each set, the splitting of *K* ($K_m$) have nothing to do with Co moment changes, even when the changes exist. It indicates that the electronic state densities of three sites change at this temperature. With further decreasing *T*, the $K_m$ values exhibit sharp decreases at $T_{c2}$, but remain finite at low temperature ($T = 5$ K). The center of gravity $K_{mt}$ of all the α, $\beta_1$ and $\beta_2$ sites estimated by taking their weighted average is also plotted by open squares, where the integrated intensities of the spectra are used as the weight. Its *T*-dependence is similar to that of χ for *H* || *c*. The *T*-dependence of $K_m$ obtained here for each distinct site represents the *T*-dependence of χ of the corresponding site.

In Fig. 10, $1/T_1T$ values of Na$_{0.5}$CoO$_2$ shown in Fig. 4 are replotted with the logarithmic vertical scale to show the characteristics of the low *T* behavior. In the inset, the similar figure is also shown for K$_{0.5}$CoO$_2$. We can find the very rapid decrease of $1/T_1T$ for both systems with decreasing *T* without particular anomalous behavior reported in ref. 17 in the low *T* region, though its increase with decreasing *T* possibly due to the randomness effects is observed in the lowest *T* region studied here.

## 4. Discussion

The internal fields and $1/T_1T$ observed at Co1 and Co2 sites of Na$_{0.5}$CoO$_2$ and K$_{0.5}$CoO$_2$ are shown in Figs. 3 and 4, respectively. The behaviors of these quantities of K$_{0.5}$CoO$_2$ are very similar at $T_{c1}$ to those of Na$_{0.5}$CoO$_2$, indicating that the magnetic structure in the *T* region between $T_{c1}$ and $T_{c2}$ is similar to that of Na$_{0.5}$CoO$_2$. In contrast to this fact, the peak found at $T_{c2}$ in the $1/T_1T$-*T* curves observed for both Co1 and Co2 sites of K$_{0.5}$CoO$_2$ is much more significant than that of Na$_{0.5}$CoO$_2$. It may indicate that the magnetic nature of the transition is more prominent in K$_{0.5}$CoO$_2$ than in Na$_{0.5}$CoO$_2$.

At $T_{c2}$, the NMR peak from Co2 of Na$_{0.5}$CoO$_2$ exhibits the splitting with decreasing *T*. In K$_{0.5}$CoO$_2$, the peak does not exhibit any tendency of such the splitting. Just the very weak anomaly of the $H_{int}$-$T$ curve exists.

At low $T$, the ratios of $H_{int}$ values at Co1 and Co2 sites are 3:1 and 2:1 for $Na_{0.5}CoO_2$ and $K_{0.5}CoO_2$, respectively. The ratios of $1/T_1T$ values of Co1 and Co2 sites at 200 K ($> T_{c1}$) are also found to be 3:1 and 2:1 for these systems, respectively. Because $H_{int}$ at $T < T_{c1}$ is determined by the static moment, while $1/T_1T$ at $T > T_{c1}$ is determined by the spectral weights of dynamical magnetic fluctuations at each kind of Co sites, they depend on the Co valences. Describing the Co1 and Co2 valences of $Na_{0.5}CoO_2$ as 3.5+δ and 3.5-δ, respectively, we represent the magnitude of δ by the difference of $\nu_Q$ values, $\Delta \nu_Q$ (= $\nu_Q^{Co1}$ - $\nu_Q^{Co2}$), between these two Co sites: Adopting a linear relationship between δ and $\nu_Q$, $\nu_Q$ = 3.43 + 5.51δ ($\nu_Q$ = 3.43 - 5.51δ) for Co1 (Co2), we roughly estimate the δ value to be ~ 0.1 for $Na_{0.5}CoO_2$. (Note that $\nu_Q$ of $NaCoO_2$, in which all Co ions have the valence of +3, is ~ 0.67 MHz,[18] and that the $\nu_Q$ values of Co1 and Co2 of $Na_{0.5}CoO_2$ are 4.1 MHz and 2.8 MHz, respectively.)

The $\Delta \nu_Q$ values of $Na_{0.5}CoO_2$ and $K_{0.5}CoO_2$ can be estimated to be 1.3 MHz and 0.7MHz, respectively, by using the $\nu_Q$ values already shown above. Because $\Delta \nu_Q$ is larger for the former system than that of the latter, the δ value is also larger for the former than for the latter. Then, we can naturally understand that the differences of $H_{int}$ and $1/T_1T$ between Co1 and Co2 sites or their ratios are expected to be larger for $Na_{0.5}CoO_2$ than for $K_{0.5}CoO_2$.

Results of the $^{23}$Na-NMR measurements have been shown in Figs. 5-9. Because the integrated intensity of the $^{23}$Na-NMR spectra of $\beta_2$ sites is rather small as compared with that of $\beta_1$ sites, we just consider that all Co atoms at B sites above $T_{c1}$ are at $\beta_1$ sites below $T_{c1}$. Assuming that the $\nu_Q$ values at Na1 and Na2 sites are mainly determined by the Co valences, we have just tried to see the relative magnitudes of the $\nu_Q$ values at these sites by the point-charge-model, and found that for δ = 0.1, $\nu_Q(Na1)/\nu_Q(Na2)$ = 1.09. (Note that the absolute magnitude of $\nu_Q$ obtained by this simple model, which does not consider the spreading of the electrons, is not so reliable, as is widely known.) The fact that this ratio is larger than unity, is consistent with the above tentative assignment that the sites α and $\beta_1$ correspond to the Na1 and Na2 sites, respectively (see Fig. 5).

The observed splitting of the center peak of α sites for $H \parallel c$ has been found to be 166×2 G. Because the dipole field is, as stated above, rather small (~ 20 G) at Na1 sites, the splitting may be understood by the existence of the transferred hyperfine field, which can become nonzero only when the $z$ coordinates of Na sites have a deviation from the midpoints between two adjacent Co layers. Actually, the deviation has been found to exist by our neutron diffraction study, as stated above.[16] The transferred hyperfine fields are created by the moments of Co2 sites. Because the transferred hyperfine field parallel to $c$ at Na2 sites can be found to be smaller than at Na1 sites, the dipole field at Na2 sites predominantly contributes to $H_{int}^c$. (Note that $H_{int}^c$ at $\beta_1$ sites or Na2 sites was found to be ~ 691 G.)

In order to find the shift magnitude of the center of gravity of the center transition lines split into two at $T_{c1}$, we have defined the mean Knight shifts $K_m$ of $^{23}$Na as the average of the shifts of two split peaks. From the magnetic structure in Fig. 7, we think that the $K_m$ values at α (Na1) and $\beta_1$ (Na2) sites are primarily determined by the spin susceptibility of Co2 and Co1 sites, respectively. In Fig. 9, we plot the $T$-dependence of $K_m$ of α, $\beta_1$ and $\beta_2$ sites. In the figure, positions of the center of gravity $K_{mt}$ of the peaks from all the sites are also shown by the open squares. The deviations of $K_m$ from $K_{mt}$ deduced from the uniform susceptibility data χ ($H \parallel c$) (shown in the figure by the broken line) is of the order of ~ 0.003 % (~ 1 G for the external field $H$ ~ 4.1 T) at $T$ between $T_{c1}$ and $T_{c2}$. Below $T_{c2}$, the deviation is larger.

At $T_{c1}$, $K_m$ of α sites deviates upward from the broken line and that of $\beta_1$ site decreases. These behaviors may be understood by considering two types of Co moments below $T_{c1}$. One corresponds to antiferromagnetically ordered electrons at Co1 sites with the smaller spin susceptibility and the larger moment, and the other corresponds to itinerant electrons at Co2 sites with the larger spin susceptibility and the smaller moment. Then, α ($\beta_1$) sites with the upward (downward) shift can be assigned to correspond to Na1 (Na2) sites located as shown in Fig. 7. This assignment is also consistent with the above result.

At $T_{c2}$, the $K_m$ values of both Na1 and Na2 sites decrease with decreasing $T$, indicating that the spin susceptibilities of both Co1 and Co2 sites exhibit a reduction as the transition at $T_{c2}$ takes place. We already know that the spin susceptibility of $K_{0.5}CoO_2$ also exhibits a rapid decrease at $T_{c2}$ with decreasing $T$.[14] $^{59}$Co-$1/T_1T$-$T$ curve of $K_{0.5}CoO_2$ exhibits a significant peak at the temperature. In $Na_{0.5}CoO_2$, the anomaly is also significant. These results suggest the occurrence of a spin-density-wave (SDW) transition, as was discussed in ref. 15. However, the anomalies of $H_{int}$ at both Co sites are very small at $T_{c2}$ (see Fig. 3). Therefore, the transition might be considered not to be magnetic one but to be due to orbital or charge ordering. If the orbital ordering is relevant, we have to understand why the $T$-dependence of $\nu_Q$ of all Co and Na sites are not significant with varying $T$ through $T_{c2}$. The decrease of $K_m$ at both α and $\beta_1$ sites, seem to indicate that the low temperature phase is in an insulating state. At low temperatures, the $K_m$ values of all Na sites remain finite, indicating that the spin susceptibilities of these sites do not vanish even at low temperatures. This might seem not to be consistent with the results of the electric specific heat measurements of $Na_{0.5}CoO_2$ (γ ~ 0).[3] However, the finite values of $K_m$ at 5 K may be explained by the magnetic moment induced along the direction of the external field.

**5. Summary**
Results of $^{59}$Co-NMR measurements on $Na_{0.5}CoO_2$ and $K_{0.5}CoO_2$ have been presented. Although the observed behaviors of the two systems are rather similar, some differences have been found: At $T_{c2}$, the $1/T_1T$-$T$ curves obtained for $K_{0.5}CoO_2$ at both Co1 and Co2 sites exhibit sharp peaks, suggesting that the transition of $K_{0.5}CoO_2$ has the stronger magnetic nature than that of $Na_{0.5}CoO_2$. The very small splitting of $H_{int}$ below $T_{c2}$ at Co2 sites of $Na_{0.5}CoO_2$ is completely absent in $K_{0.5}CoO_2$, which makes it unclear whether the splitting commonly characterizes the transitions at $T_{c2}$. We have described the valences of Co1 and Co2 ions of $Na_{0.5}CoO_2$ as $3.5+\delta$ and $3.5-\delta$ with $\delta \sim 0.1$, respectively.

$^{23}$Na-NMR measurements for $Na_{0.5}CoO_2$ under the fields parallel and perpendicular to the $c$-axis have also been carried out. Above $T_{c1}$, the spectra consist of the contributions from the two crystallographically distinct Na sites, and below $T_{c1}$, the lines from these two Na sites shift to both the lower and higher frequency sides due to the internal magnetic field created by the ordered Co moments. The detailed spectra can be explained by the magnetic structure with the magnetic moments parallel and perpendicular to $c$, reported in ref. 3. We have shown that the local electronic densities of states at all Na sites change at the transition temperatures.

To understand the detailed mechanism of the transition at $T_{c2}$, it is important to consider the following facts: (1) The anomalies of the internal fields $H_{int}$ observed at $T_{c2}$ are very small. (2) The splitting of the zero-field NMR spectra observed at $T_{c2}$ with decreasing $T$ for Co2 sites of $Na_{0.5}CoO_2$ does not exist for $K_{0.5}CoO_2$. (3) The $\nu_Q$ values of all Co and Na sites do not exhibit significant $T$-dependence with varying $T$ at around $T_{c2}$.

Acknowledgments –The work is supported by Grants-in-Aid for Scientific Research from the Japan Society for the Promotion of Science (JSPS) and by Grants-in-Aid on priority area from the Ministry of Education, Culture, Sports, Science and Technology.

After submitting this paper, we have realized a work by Ning et al. [arXiv:0711.4023v1], which points out that the $\mathbf{H}_{int}$ at Co2 sites are not parallel to the $c$-axis, and that the observed spectra should be analyzed by the exact calculation instead of the present second order perturbation method. Actually, because the presence of the dipole fields from the ordered Co moments induce the deviation of the direction of $\mathbf{H}_{int}$ from the $c$-axis and because the condition $\nu_Q \ll$ (the Zeeman energy) is not well satisfied, we have tried to see how much the present results are changed by considering these facts, and found followings. (1) The dipole field at Co2 sites is roughly estimated to be ~700 G (~ 450 G) for $Na_{0.5}CoO_2$ ($K_{0.5}CoO_2$), which induces the deviation of $\mathbf{H}_{int}$ from the $c$-axis toward the $a$-axis about 6° (4°), which should be compared with 9° reported by Ning et al. for $Na_{0.5}CoO_2$. (2) For $Na_{0.5}CoO_2$, the $H_{int}$ values corresponding to the two split peaks of the Co2 spectra are different from those of the perturbation calculations reported above by +0.1 % and +1.9 % at 9.8 K for the lower and higher frequency peaks, respectively, as shown in the inset of Fig. 3(a) by the solid squares. The difference between the values of $H_{int}$ of Co1 sites is +0.3 % as shown by the open squares at 9.8 K. Detailed parameters determined at 9.8 K by the exact diagonalizations are: $\eta = 0.3$, $\nu_Q$ (= $\nu_{zz}$) = 2.79 MHz, $H_{int}$ = 0.706 T, $\theta$ = 6.55° and $\varphi$ = 90° ($\nu_{zz} > \nu_{xx} > \nu_{yy}$; $z \parallel c^{3)}$) for Co2(a), $\eta = 0.3$, $\nu_Q = 2.885$ MHz, $H_{int}$ = 0.672 T, $\theta$ = 7.8° and $\varphi$ = 90° ($z \parallel c^{3)}$) for Co2(a). For Co1, $\eta = 0.4$, $\nu_Q = 4.03$ MHz, $H_{int}$ = 2.084 T, $\theta$ = 90° and $\varphi$ = 0° ($z \parallel c^{3)}$). These parameters do not have significant difference from the values obtained by the perturbation calculation described above. For $K_{0.5}CoO_2$, $H_{int}$ changes +3.6 % and -0.3 % for Co2 and Co1 sites at 4.2 K, as shown in the inset of Fig. 3(b) by the solid and open squares, respectively. (For $K_{0.5}CoO_2$, the peak does not split.) The detailed parameters determined by the exact diagonalization to reproduce the spectra observed at 4.2 K are: $\eta = 0.28$, $\nu_Q = 3.125$ MHz, $H_{int}$ = 0.603 T, $\theta$ = 9.5° and $\varphi$ = 90° ($z \parallel c^{3)}$) for Co2, and $\eta = 0.23$, $\nu_Q = 3.83$ MHz, $H_{int}$=1.286 T, $\theta$ = 90° and $\varphi$ = 0° ($z \parallel c^{3)}$) for Co1. These parameters are rather different from those given by the perturbation calculation, and the lowest frequency peak from Co1 sites shown in ref. 14, should be assigned to the one at ~ 7.5 MHz instead of the one at 9 MHz. (3) The main fraction of the $1/2 \leftrightarrow -1/2$ transition of Co2 site (Co2(a) and Co2(b) sites) is found in the lines indicated by open squares in the top panel of Figs. 1(b) and 1(c). It is consistent with the present assignments by the second order perturbation method and different from the results of Ning et al.

Although detailed parameters and the parts of the assignments should be revised, as described above, the essential results of the paper do not change.

We stress that the peaks from Co2 sites of $K_{0.5}CoO_2$ shown in ref. 14 do not exhibit such the splitting as observed in $Na_{0.5}CoO_2$ below $T_{c2}$, even though the resistivity of the system begins to increase significantly at $T_{c2}$ with decreasing $T$ in a similar way to the case of $Na_{0.5}CoO_2$. Therefore, we suspect if the splitting of the peaks can be the evidence, as pointed out by Ning et al, for the charge ordering.

**Figure caption**

Fig. 1 (a) $T$-dependence of the magnetic susceptibilities $\chi$ measured for a single crystal under applied fields $H$ (= 1 T) perpendicular (open circles) and parallel (open squares) to $c$ and the electrical resistivity $\rho$ (close circles) of a polycrystalline sample of $Na_{0.5}CoO_2$. (b) The zero-field NMR spectra of $Na_{0.5}CoO_2$ at 9.8 K ($< T_{c2}$)[3] are shown in the bottom panel. The spectra calculated by the perturbation of the $eqQ$ interaction up to the second order (middle panel) and by the exact diagonalization (top panel) are shown for Co1 (black lines) and Co2 (grey lines). The main fraction of the 1/2 ↔ -1/2 transition lines of Co1 and Co2 sites are indicated by the open circles and open squares. See text for detailed parameters. (c) The zero-field NMR spectra of $K_{0.5}CoO_2$ at 4.2 K ($< T_{c2}$)[14] are shown in the bottom panel. Results of the calculations are also shown in the top and middle panels similarly in (b). See text for detailed parameters.

Fig. 2 The zero-field $^{59}$Co NMR profiles observed at Co2 sites are shown for $Na_{0.5}CoO_2$ and $K_{0.5}CoO_2$ at the first-third satellite positions. Note that no indication of the peak splitting can be found for $K_{0.5}CoO_2$. No broadenings of the lines have not been observed with decreasing $T$ through $T_{c2}$.

Fig. 3 (color online) (a) Internal magnetic fields $H_{int}$ at Co1 (open circles) and Co2 (small dots) sites estimated by the perturbation calculations are plotted against $T$ for $Na_{0.5}CoO_2$. These fields appear at $T_{c1}$ and increase with decreasing $T$. At $T_{c2}$, very tiny splitting exists for Co2 sites, as can be found in the inset, where 3.04×(absolute values of $H_{int}$ (∥ $c$)) at Co2 sites are plotted with the absolute values of $H_{int}$ (∥ $ab$-plane) at Co1 sites. Open and solid squares indicate the values of $H_{int}$ (∥ $ab$-plane) at Co1 sites and 3.04×(absolute values of $H_{int}$ (∥ $c$)) at Co2 sites obtained by the exact diagonalization. See text for details. (b) Similar figure to (a) is shown for $K_{0.5}CoO_2$. The data were obtained by the perturbation calculations. The inset also shows the similar figure to that in (a), but we use the values of 2.17×(absolute values of $H_{int}$ (∥ $c$)) instead of 3.04×(absolute values of $H_{int}$ (∥ $c$)). We do not see any indication of the peak splitting of Co2 sites even below $T_{c2}$.

Fig. 4 The longitudinal relaxation rates divided by $T$, $1/T_1T$ of $^{59}$Co nuclei, observed for $K_{0.5}CoO_2$, are shown together with those of $Na_{0.5}CoO_2$. As in the case of $Na_{0.5}CoO_2$, $1/T_1T$-$T$ curves of both Co1 and Co2 sites exhibit sharp peaks at $T_{c1}$ in $K_{0.5}CoO_2$. At $T_{c2}$, $1/T_1T$ also exhibits sharp peaks for both Co sites of $K_{0.5}CoO_2$.

Fig. 5 $^{23}$Na-NMR spin-echo intensities times $T$ measured with fixed $H$ (~ 4.1 T) parallel to $c$ for $c$-axis aligned crystallites of $Na_{0.5}CoO_2$ are plotted against $f$-$^{23}\gamma H$ at four temperatures, where $f$ is the NMR frequency and $^{23}\gamma$ is gyromagnetic ratio of $^{23}$Na nuclei.

Fig. 6 Temperature dependences of the $c$-axis component of the internal magnetic field $H_{int}^c$ at Na sites are shown. These values were estimated by taking the half distance between the center-line peaks split due to the internal field created by the Co ordering.

Fig. 7 The magnitude and direction of the dipole fields at Na2 sites roughly estimated by using the magnetic structures of $Na_{0.5}CoO_2$, where only the nearest neighbor Co atoms are considered. At Na1 sites, the magnitude of the dipole field is negligibly small (~ 20 G (∥ $c$) and ~ 5 G (⊥ $c$)), even when the contributions from the nearest and next nearest neighbor Co atoms are considered.

Fig. 8 $^{23}$Na-NMR spin-echo intensities times $T$ taken at two temperatures for the $c$-axis aligned crystallites of $Na_{0.5}CoO_2$ are shown against the difference between the NMR frequency and $^{23}\gamma H$. The data at 98.8 K (closed circles) were taken by changing the NMR frequency stepwise with the applied field $H$ (∥ $ab$-plane) being fixed at ~ 4.1 T, and the data at 8.1 K (open circles) were taken by changing the field $H$ (∥ $ab$-plane) stepwise with the frequency $f$ being fixed. The inset shows the central transition line. The solid lines are obtained by the calculations, for which following parameters are used. For A sites, $\nu_Q$ = 1.32 MHz and $\eta$ = 0.62 and for B sites, $\nu_Q$ = 1.26 MHz and $\eta$ = 0.52 at 98.8 K. At 8.1 K ($< T_{c2}$), the in-plane component of the internal fields $H_{int}^{ab}$ ~ 40 G (~ 120 G) is also considered in the calculation at $\alpha$ ($\beta_1$) sites, where $\nu_Q$ = 1.30 (1.25) MHz and $\eta$ = 0.62 (0.47) are used.

Fig. 9 Temperature dependences of the $^{23}$Na-NMR Knight shift $K$ ($K_m$, $K_{mt}$) of $Na_{0.5}CoO_2$ measured under the magnetic field parallel to $c$ are shown together with the $T$-dependence of $\chi$. These values below $T_{c1}$ were estimated by using the mean position of central transition lines (see text for details).

Fig. 10 The $1/T_1T$ values of $Na_{0.5}CoO_2$ are plotted at two Co sites with the logarithmic vertical scale against $T$. Inset shows the $T$-dependence of $1/T_1T$ for $K_{0.5}CoO_2$ at two Co sites.

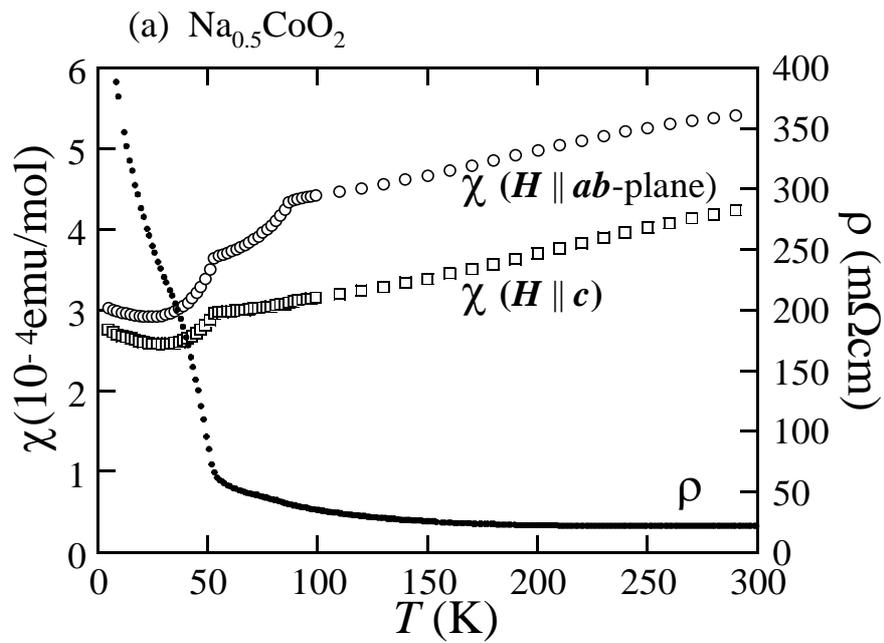

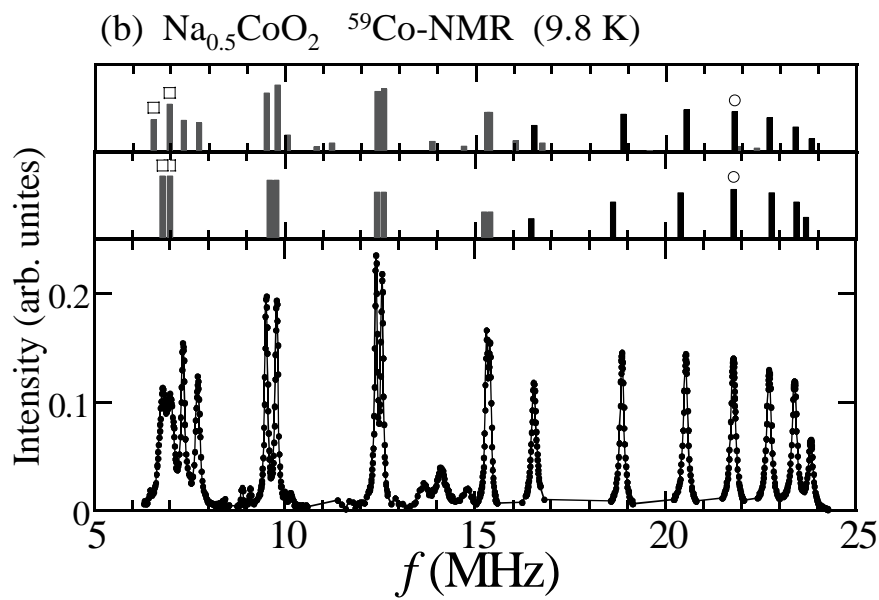

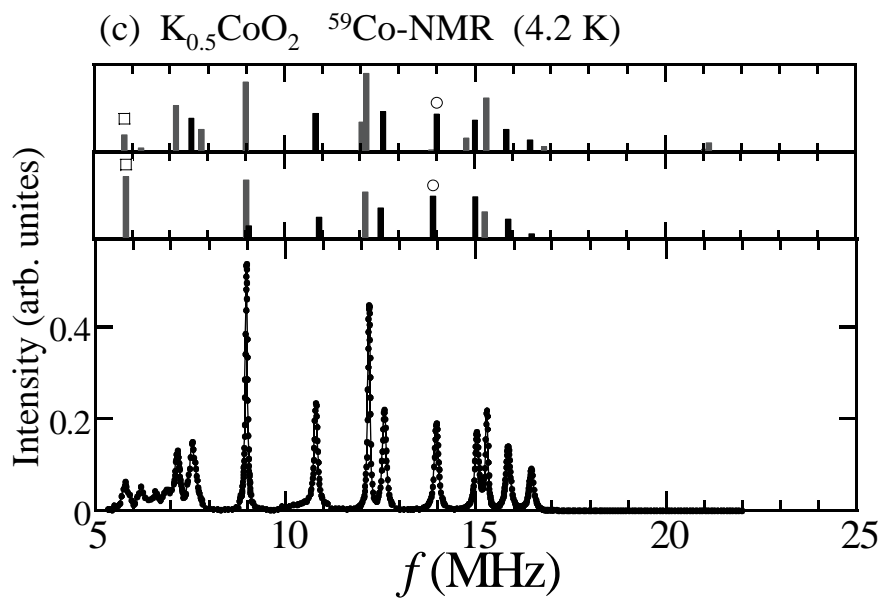

Fig. 1(a), 1(b), 1(c)     M. Yokoi *et al.*

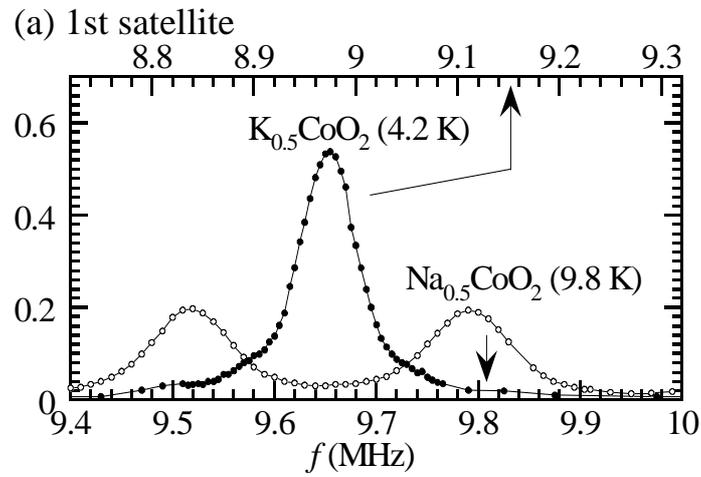

(a) 1st satellite

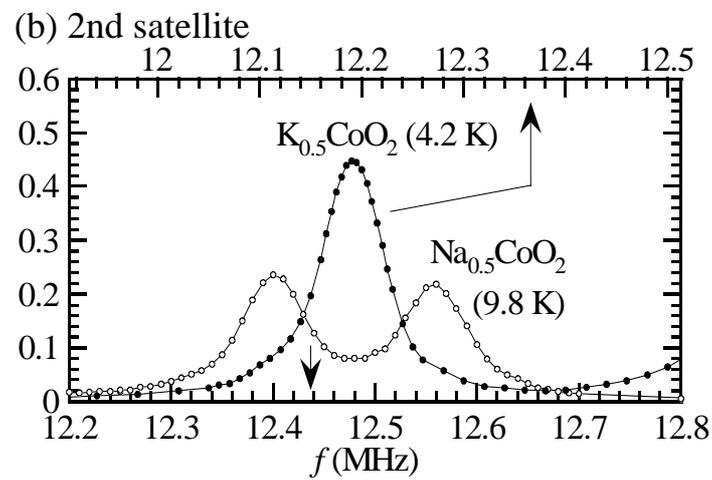

(b) 2nd satellite

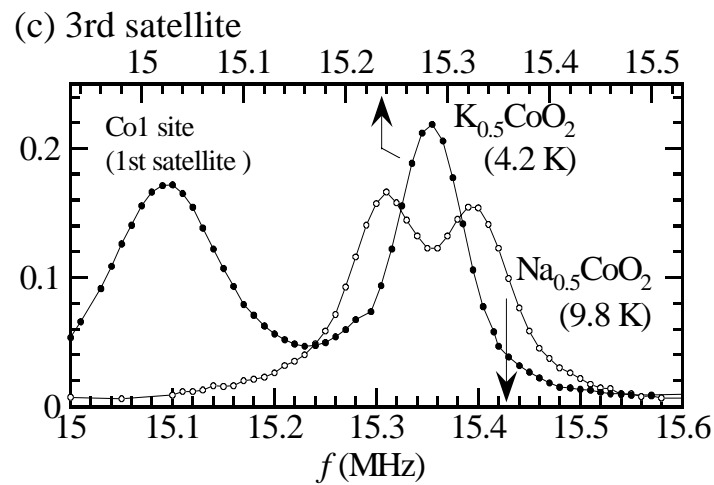

(c) 3rd satellite

Fig. 2(a), 2(b), 2(c)
M. Yokoi *et al.*

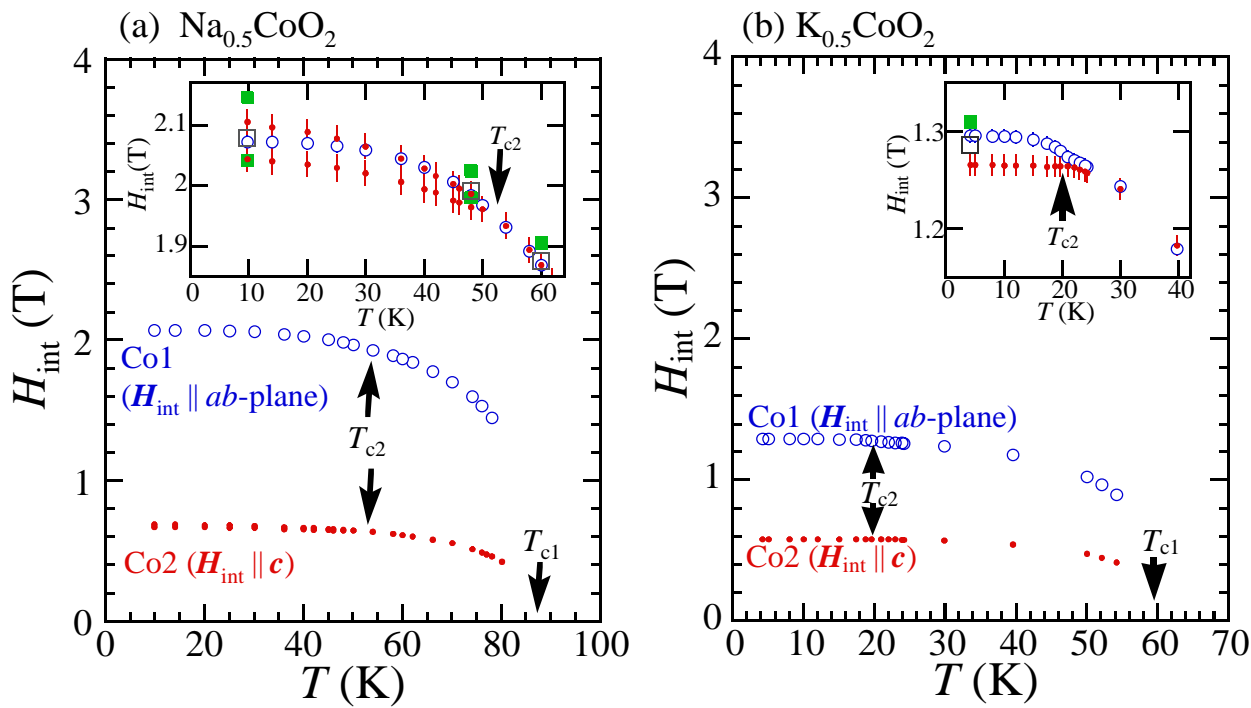

Fig. 3(a), 3(b)

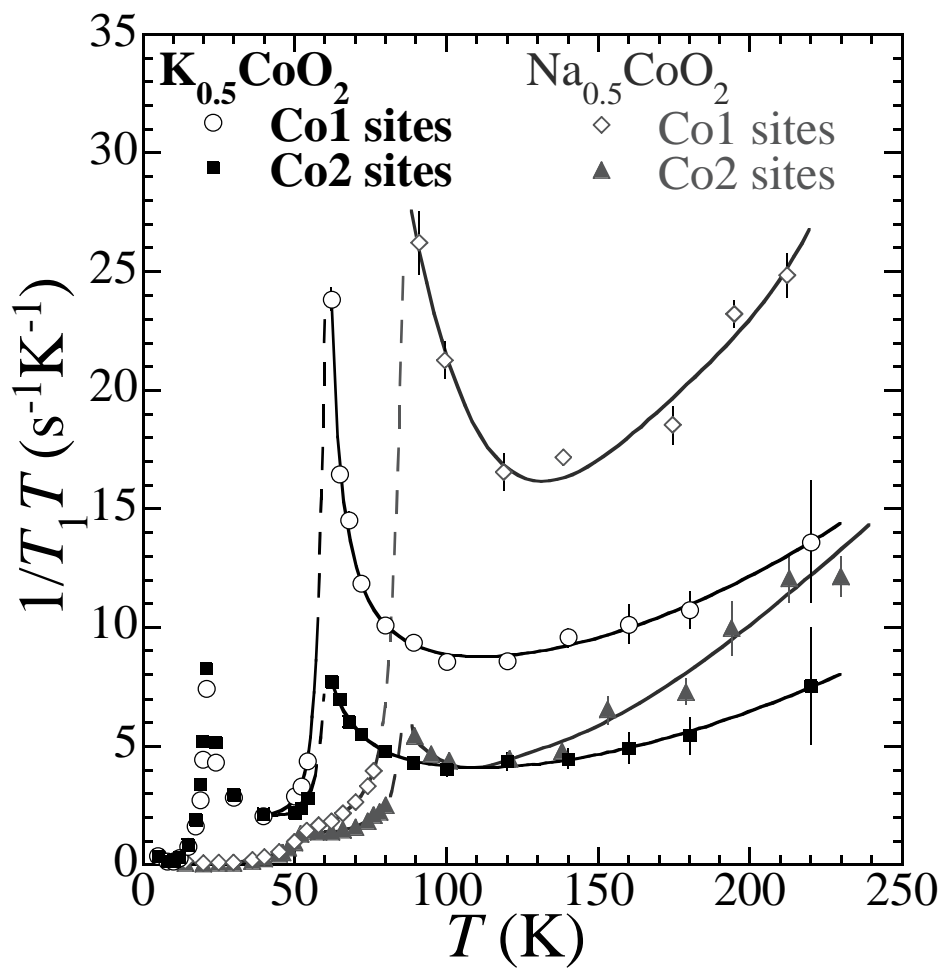

Fig. 4     M. Yokoi *et al.*

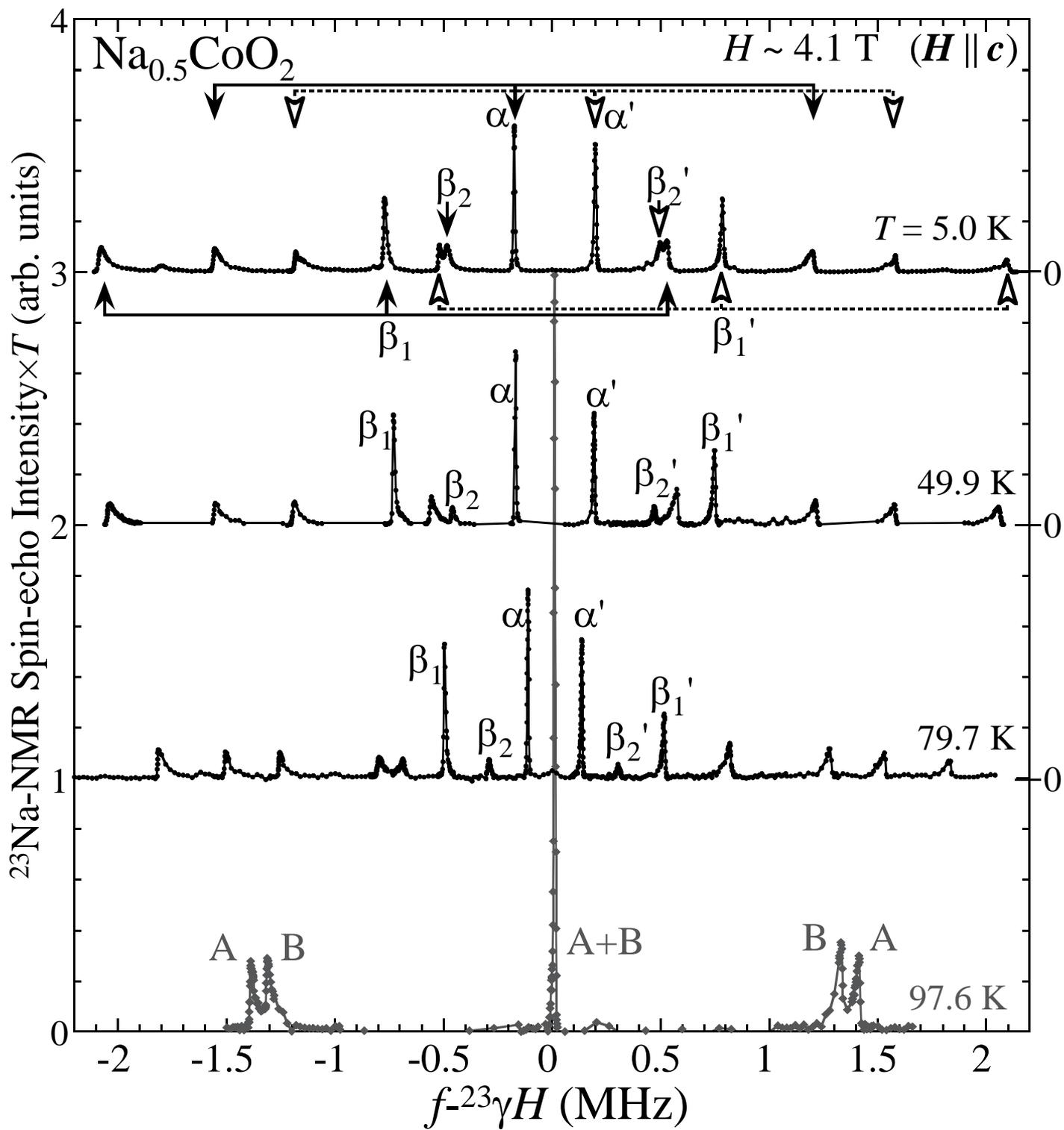

Fig. 5

M. Yokoi et al.

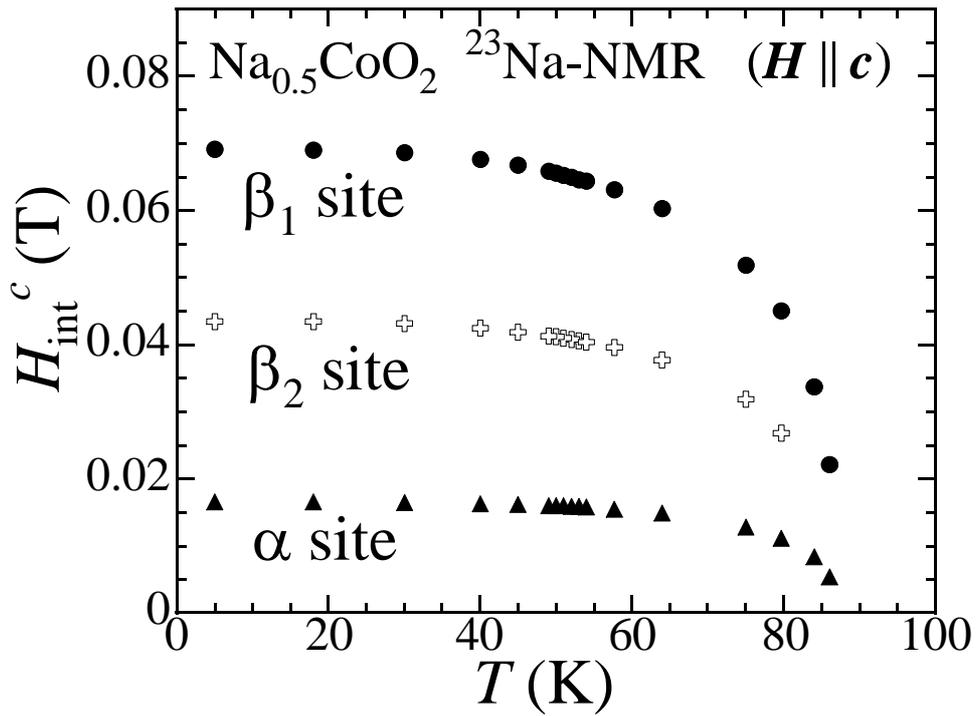

Fig. 6

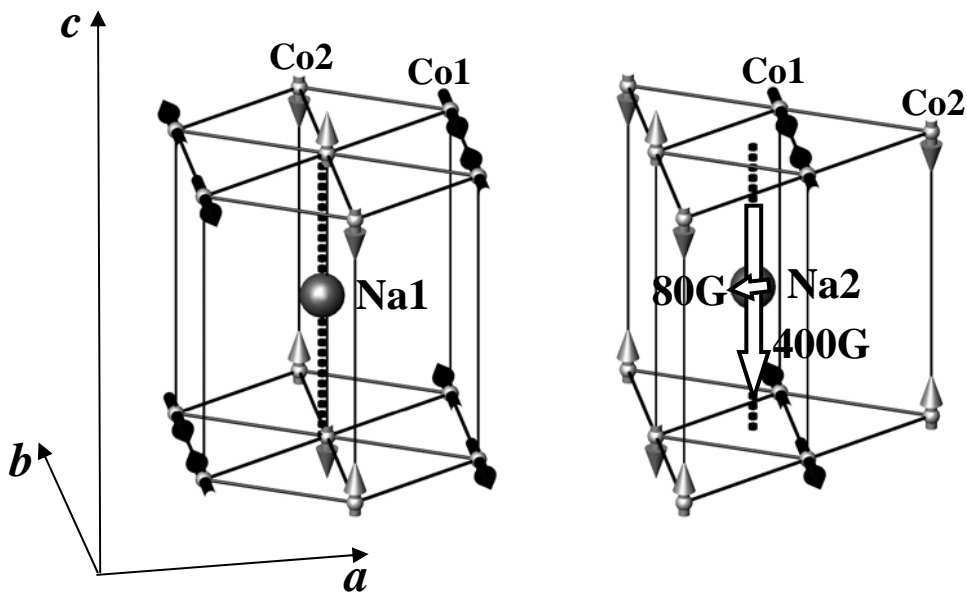

Fig. 7

M. Yokoi *et al.*

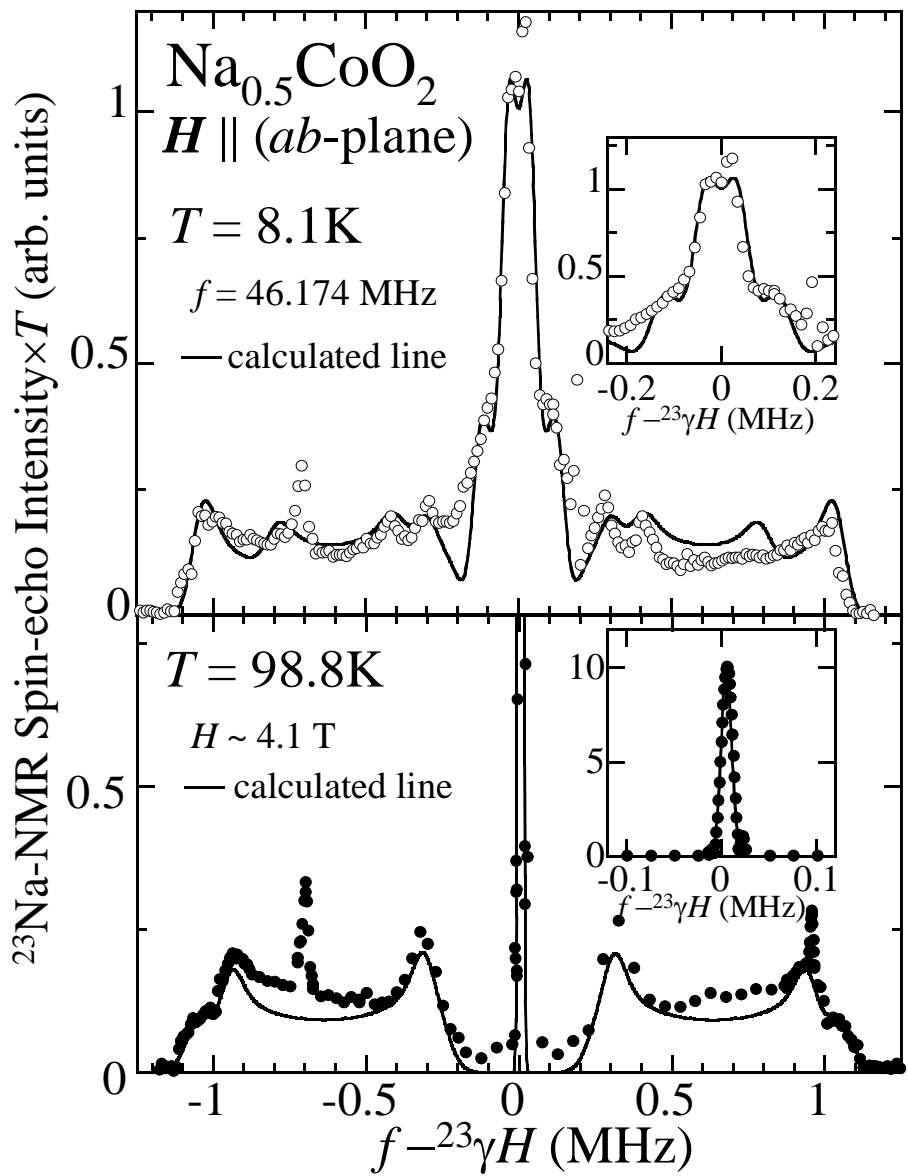

Fig. 8
M. Yokoi et al.

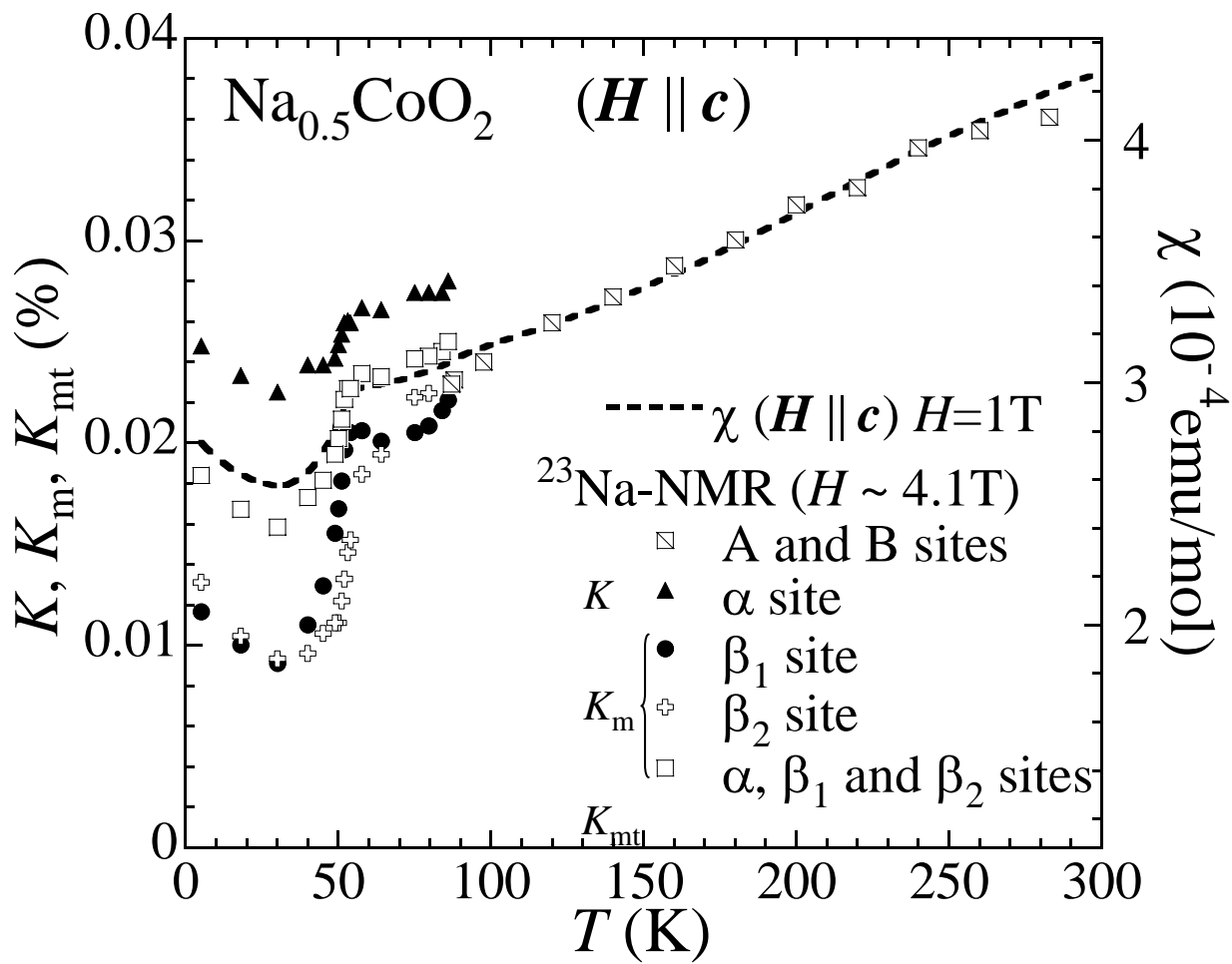

Fig. 9

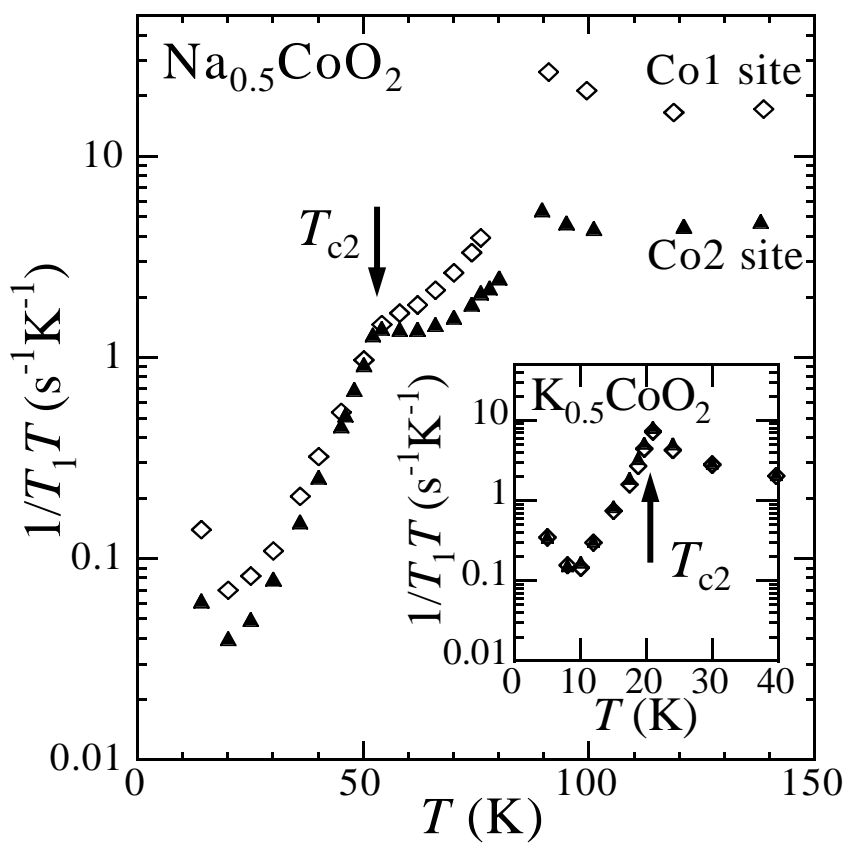

Fig. 10   M. Yokoi *et al.*